\documentclass[twocolumn,showpacs,pra,aps,amsmath]{revtex4}
\usepackage{graphicx}

\begin{document}

\title{Electron-hydrogen scattering in 
Faddeev-Merkuriev integral equation approach}

\author{Z.\ Papp${}^{1}$ and C-.Y.\ Hu${}^{2}$}
\affiliation{${}^{1}$ Institute of Nuclear Research of the
Hungarian Academy of Sciences, Debrecen, Hungary\\
${}^{2}$ Department of Physics and Astronomy, 
California State University, Long Beach, 90840, California  }
\date{\today}

\begin{abstract}
Electron-hydrogen scattering is studied in the
Faddeev-Merkuriev integral equation approach. The equations are
solved by using the Coulomb-Sturmian separable expansion technique.
We present $S$- and $P$-wave scattering and reactions cross sections up 
to the $H(n=4)$ threshold.
\end{abstract}


\maketitle

\section*{Introduction}

The scattering of electrons  on hydrogen atom is a fundamental three-body
problem in atomic physics. The long-range Coulomb interaction presents the 
major difficulty. On the other hand, it is a special kind 
of Coulomb three-body problem as it contains two identical particles.
While many studies have been carried out aiming at solving the Schr\"odinger
equation using perturbative, close-coupling, variational or direct numerical 
methods approaches along to the Faddeev equations are relatively scarce.
Here, by solving Faddeev-type integral equations,
we present a general numerical method suitable for the treatment of 
elastic and inelastic processes in three-body Coulombic systems with
two identical particles and apply the formalism for the electron-hydrogen
system.
 
For  quantum mechanical three-body systems
the Faddeev integral equations are the fundamental equations.
They possess connected kernels and therefore they are Fredholm-type
integral equations of second kind. 
 The Faddeev equations were derived for short range
interactions and if we simply plug-in a Coulomb-like potential they become
singular. The necessary modification were proposed by Merkuriev \cite{fm-book}.
In Merkuriev's approach the Coulomb interactions were split into short-range
and long-range parts. The long-range parts were included into the
,,free'' Green's operators and the Faddeev procedure were performed only with
the short-range potentials.
The corresponding modified Faddeev, or Faddeev-Merkuriev equations 
are mathematically well-behaved. They possess compact kernels even in the case
of attractive Coulombic interactions. This means that the Faddeev-Merkuriev 
equations possess all the nice properties of the original Faddeev equations.

However, the associated three-body Coulomb Green's operator is not known
explicitly. To circumvent the problem the integral equations were cast into 
differential form and the appropriate boundary conditions were derived from the
asymptotic analysis of the three-body Coulomb Green's operator. 
These modified Faddeev differential equations  were successfully solved for 
various atomic three-body problems, 
including electron-hydrogen scattering up to the $H(n=3)$ threshold \cite{kwh}.

A characteristic property of the atomic three body systems is that, 
due to attractive Coulomb interactions, they have infinitely many 
two-body channels. If the total energy of the system increases more and more
channels open up. The differential equation approach needs
boundary conditions for each channels, and becomes intractable if the energy
increases beyond a limit.
Integral equations do not need boundary conditions, this information is 
incorporated in the Green's operators. They need initial conditions, which are
much simpler. Therefore an integral equation approach to the three-body Coulomb
problem would be very useful, it could provide an unified description of the
scattering and reactions processes for all energies.

In the past few years we have developed a new
approach to the three-body Coulomb problem.
Faddeev-type integral equations were solved by using the Coulomb-Sturmian
separable expansion method. The approach was developed first
for solving the nuclear three-body scattering problem with repulsive 
Coulomb interactions \cite{pzsc}, which has been 
adapted recently for atomic systems with attractive Coulomb 
interactions \cite{phhky}. The basic concept in this method is a 
,,three-potential'' picture, where the $S$ matrix is given in three terms.
In this approach we  solve the Faddeev-Merkuriev integral equations such that
the associated three-body Coulomb Green's operator is calculated by an
independent Lippmann-Schwinger-type integral equation.
This Lippmann-Schwinger integral equation contains the channel-distorted 
Coulomb Green's operator, which can be calculated as a contour integral of 
two-body Coulomb Green's operators.
The method were tested in positron-hydrogen scattering for energies up to
the $H(n=2)-Ps(n=2)$ gap  \cite{phhky}, and good agreements with
the configuration-space solution of the Faddeev-Merkuriev equations
were found.
 
In this paper we apply this formalism for the electron-hydrogen scattering
problem. In Sec.\ I we briefly describe the Faddeev-Merkuriev integral 
equations, the details are given in 
Ref.\ \cite{phhky}. However, the fact that in the electron-hydrogen system we
have to deal with identical particles requires some additional considerations:
the symmetry simplifies the numerical procedure.
In Sec.\ II the integral equations are solved by the Coulomb-Sturmian 
separable expansion method.
In Sec.\ III we show  some test calculations up to the $H(n=4)$ threshold
with total angular momenta $L=0$ and $L=1$. Finally,
we draw some conclusions.

\section{Faddeev-Merkuriev integral equations for the $e^{-}+H$ system}

In the $e^{-}+H$ system the two electrons are identical. 
Let us denote them
by $1$ and $2$, and the non-identical proton by $3$.
The  Hamiltonian is given by
\begin{equation}
H=H^0 + v_1^C + v_2^C + v_3^C,
\label{H}
\end{equation}
where $H^0$ is the three-body kinetic energy 
operator and $v_\alpha^C$ denotes the Coulomb
interaction in the subsystem $\alpha$. 
We use the usual configuration-space Jacobi coordinates
 $x_\alpha$ and $y_\alpha$, where $x_\alpha$ is the coordinate
between the pair $(\beta,\gamma)$ and $y_\alpha$ is the
coordinate between the particle $\alpha$ and the center of mass
of the pair $(\beta,\gamma)$.
Thus the potential $v_\alpha^C$, the interaction of the
pair $(\beta,\gamma)$, appears as $v_\alpha^C (x_\alpha)$.
The  Hamiltonian (\ref{H}) is defined in the three-body 
Hilbert space. So, the two-body potential operators are formally
embedded in the three-body Hilbert space,
\begin{equation}
v^C = v^C (x) {\bf 1}_{y},
\label{pot0}
\end{equation}
where ${\bf 1}_{y}$ is a unit operator in the two-body Hilbert space associated
with the $y$ coordinate. 

The role of a Coulomb potential in a three-body
system is twofold. In one hand, it acts like a long-range potential
since it modifies the asymptotic motion. On the other hand, however,
it acts like a short-range potential, since it correlates strongly the 
particles and may support bound states.
Merkuriev introduced a separation of the three-body
configuration space into different
asymptotic regions \cite{fm-book}. 
The two-body asymptotic region $\Omega$ is
defined as a part of the three-body configuration space where
the conditions
\begin{equation}
|x| <  x_0 ( 1  + |y|/ y_0)^{1/\nu},
\label{oma}
\end{equation}
with $x_0, y_0 >0$ and $\nu > 2$, are satisfied.
Merkuriev proposed to split the Coulomb interaction  in 
the three-body configuration space into
short-range and long-range terms 
\begin{equation}
v^C =v^{(s)} +v^{(l)} ,
\label{pot}
\end{equation}
where the superscripts
$s$ and $l$ indicate the short- and long-range
attributes, respectively. 
The splitting is carried out with the help of a splitting function $\zeta$,
\begin{subequations}
\begin{eqnarray}
v^{(s)} (x,y) & = & v^C(x) \zeta (x,y),
\\
v^{(l)} (x,y) & = & v^C(x) \left[1- \zeta (x,y) \right].
\label{potl}
\end{eqnarray}
\end{subequations}
The function  $\zeta$  vanishes asymptotically within the three-body sector,
where $x\sim y \to \infty$, and approaches one in the two-body asymptotic region
$\Omega$, where $x<< y \to \infty$. 
Consequently in the three-body sector $v^{(s)}$
vanishes and $v^{(l)}$ approaches $v^{C}$.
In practice usually the functional form
\begin{equation}
\zeta (x,y) =  
2/\left\{1+ \exp \left[ {(x/x_0)^\nu}/{(1+y/y_0)} \right] \right\},
\label{oma1}
\end{equation}
is used.

In the Hamiltonian (\ref{H}) the Coulomb potential $v_3^C$, 
the interaction between
the two electrons, is repulsive, and does not support bound states. 
Consequently, there are no two-body channels associated with this 
fragmentation. Therefore
the entire $v_3^C$ can be considered as long-range potential. Then,
the long-range Hamiltonian is defined as
\begin{equation}
H^{(l)} = H^0 + v_1^{(l)}+ v_2^{(l)}+ v_3^{C},
\label{hl}
\end{equation}
and the three-body Hamiltonian takes the form
\begin{equation}
H = H^{(l)} + v_1^{(s)}+ v_2^{(s)}.
\label{hll}
\end{equation}
So, the Hamiltonian (\ref{hll}) appears formally
as a three-body Hamiltonian with two short-range potentials.
The bound-state wave function $|\Psi \rangle$ satisfies the homogeneous
Lippmann-Schwinger integral equation
\begin{equation}
|\Psi\rangle= G^{(l)} \left[ v_1^{(s)}+ v_2^{(s)} \right] |\Psi\rangle =
 G^{(l)}  v_1^{(s)} |\Psi\rangle +  G^{(l)}  v_2^{(s)} |\Psi\rangle,
\label{Psi}
\end{equation}
where $G^{(l)}(z)=(z-H^{(l)})^{-1}$ is the resolvent 
operator of $H^{(l)}$.
This induce, in the spirit of the Faddeev procedure, the splitting of the  
wave function $|\Psi\rangle$ into two components
\begin{equation}
|\Psi\rangle=|\psi_1\rangle  +|\psi_2\rangle,
\label{psi}
\end{equation}
where the components are defined by
\begin{equation}
|\psi_\alpha \rangle= G^{(l)} v_\alpha^{(s)} |\Psi\rangle,
\label{psidef}
\end{equation}
with $\alpha=1,2$. The components satisfy the
set of two-component Faddeev-Merkuriev integral equations
\begin{subequations}\label{fm2comp}
\begin{eqnarray}
| \psi_1 \rangle &= | \Phi_1^{(l)} \rangle + & G_1^{(l)} v_1^{(s)} 
| \psi_2  \rangle  \label{fm2c1}\\
| \psi_2 \rangle &= \phantom{\ | \phi_1 \rangle + } & 
G_2^{(l)} v_2^{(s)} | \psi_1 \rangle,
\label{fm2c2}
\end{eqnarray}
\end{subequations}
where $G_\alpha^{(l)}$ is the
resolvent operator of the channel Coulomb Hamiltonian 
\begin{equation}
H_\alpha^{(l)}=H^{(l)}+v_\alpha^{(s)}
\end{equation} 
and the inhomogeneous term $|\Phi_1^{(l)}\rangle$ is an 
eigenstate of $H^{(l)}_1$.

Before going further let us examine the spectral properties of the
Hamiltonian 
\begin{equation}
H_1^{(l)}=H^{(l)}+v_1^{(s)}=H^0+v_1^C+v_2^{(l)}+v_3^C.
\end{equation}
It obviously supports infinitely many two-body channels associated with the 
bound states of the attractive Coulomb potential $v_1^C$. 
The potential $v_3^C$ is repulsive and does not have bound states.
The three-body potential $v_2^{(l)}$ is attractive and constructed such that 
$v_2^{(l)}(x_2,y_2)\to 0$ if $y_2 \to \infty$. Therefore, there are no
two-body channels associated with fragmentations $2$ and $3$, the Hamiltonian 
$H_1^{(l)}$ has only $1$-type two-body asymptotic channels. Consequently, the 
corresponding  $G_1^{(l)}$ Green's operator, acting on the $v_1^{(s)} 
| \psi_2  \rangle$ term in (\ref{fm2c1}), 
will generate only $1$-type two-body asymptotic channels in $|\psi_1\rangle$.
Similar analysis is valid also for $|\psi_2\rangle$. Thus, the Faddeev-Merkuriev
procedure results in the separation of the three-body wave function into
components such a way that each component has only one type of two-body
asymptotic channels. This is the main advantage of the Faddeev equations and,
as this analysis shows, this property remains true also for attractive 
Coulomb potentials if the Merkuriev splitting is adopted.

In the $e^- e^- p$ system  the particles $1$ and $2$, 
the two electrons, are identical and 
indistinguishable. Therefore, the Faddeev components 
$| \psi_1 \rangle$ and $| \psi_2 \rangle$, in their own natural Jacobi
coordinates, should have the same functional forms
\begin{equation}
\langle x_1 y_1 | \psi_1 \rangle = \langle x_2 y_2 | \psi_2 \rangle
= \langle x y | \psi \rangle.
\end{equation}
On the other hand, by interchanging the two electrons we have
\begin{equation}
{\mathcal P} | \psi_1 \rangle = p | \psi_2 \rangle,
\end{equation}
where the operator ${\mathcal P}$ describes the exchange of 
particles $1$ and $2$, and $p=\pm 1$ is the eigenvalue of ${\mathcal P}$.
Building this information into the formalism results 
the integral equation
\begin{equation} \label{fmp}
| \psi \rangle = | \Phi_1^{(l)} \rangle +  G_1^{(l)} v_1^{(s)} p {\mathcal P} 
| \psi \rangle,
\end{equation}
which is alone sufficient to determine $| \psi \rangle$.
We notice that so far no approximation has been made, and although 
this Faddeev-Merkuriev integral equation has only one component, yet it
gives a full account on the asymptotic and symmetry properties of the system.

\section{Coulomb-Sturmian separable expansion approach}

We solve this integral equation
by applying the Coulomb--Sturmian separable expansion approach.
This approach has been established in a series of papers for 
two- \cite{cspse2} and three-body \cite{cspse3,pzsc,phhky} 
problems with Coulomb-like potentials.
The Coulomb-Sturmian (CS) functions are defined by
\begin{equation}
\langle r|n l \rangle =\left[ \frac {n!} {(n+2l+1)!} \right]^{1/2}
(2br)^{l+1} \exp(-b r) L_n^{2l+1}(2b r),  \label{basisr}
\end{equation}
with $n$ and $l$ being the radial and
orbital angular momentum quantum numbers, respectively, and $b$ is the size
parameter of the basis.
The CS functions $\{ |n l\rangle \}$
form a biorthonormal
discrete basis in the radial two-body Hilbert space; the biorthogonal
partner defined  by 
$\langle r |\widetilde{n l}\rangle=\langle r |{n l}\rangle/r$. 

Since the three-body Hilbert space is a direct product of two-body
Hilbert spaces an appropriate basis is the bipolar basis, which
can be defined as the
angular momentum coupled direct product of the two-body bases, 
\begin{equation}
| n \nu  l \lambda \rangle_\alpha =
 | n  l \rangle_\alpha \otimes |
\nu \lambda \rangle_\alpha, \ \ \ \ (n,\nu=0,1,2,\ldots),
\label{cs3}
\end{equation}
where $| n  l \rangle_\alpha$ and $|\nu \lambda \rangle_\alpha$ are associated
with the coordinates $x_\alpha$ and $y_\alpha$, respectively.
With this basis the completeness relation
takes the form (with angular momentum summation implicitly included)
\begin{equation}
{\bf 1} =\lim\limits_{N\to\infty} \sum_{n,\nu=0}^N |
 \widetilde{n \nu l \lambda } \rangle_\alpha \;\mbox{}_\alpha\langle
{n \nu l \lambda} | =
\lim\limits_{N\to\infty} {\bf 1}^{N}_\alpha,
\end{equation}
where $\langle x y | \widetilde{ n \nu l \lambda}\rangle = 
\langle x y | { n \nu l \lambda}\rangle/(x y)$.

We make the following approximation on the 
integral equation (\ref{fmp})
\begin{equation} \label{fmpa}
| \psi \rangle = | \Phi_1^{(l)} \rangle +  G_1^{(l)}
{\bf 1}^{N}_1 v_1^{(s)} p {\mathcal P} {\bf 1}^{N}_1 | \psi \rangle,
\end{equation}
i.e.\ the operator 
$v_1^{(s)} p {\mathcal P}$ is approximated in the three-body
Hilbert space  by a separable form, viz.
\begin{eqnarray}
v_1^{(s)}p {\mathcal P}  & = & \lim_{N\to\infty} 
{\bf 1}^{N}_1 v_1^{(s)} p {\mathcal P}  {\bf 1}^{N}_1 \nonumber \\ 
& \approx & {\bf 1}^{N}_1 v_1^{(s)} p {\mathcal P} {\bf 1}^{N}_1 \nonumber \\ 
& \approx  & \sum_{n,\nu ,n', \nu'=0}^N
|\widetilde{n\nu l \lambda}\rangle_1 \; \underline{v}_1^{(s)}
\;\mbox{}_1 \langle \widetilde{n' \nu' l' \lambda'}|,  \label{sepfe}
\end{eqnarray}
where $\underline{v}_1^{(s)}=\mbox{}_1 \langle n\nu l \lambda|
v_1^{(s)} p {\mathcal P} |n' \nu' l' \lambda' \rangle_1$.
Utilizing the properties of the exchange operator ${\mathcal P}$
these matrix elements can be written in the form 
$\underline{v}_1^{(s)}= p\times (-)^{l'} \; \mbox{}_1 \langle n\nu l \lambda| 
v_1^{(s)}|n' \nu' l' \lambda' \rangle_2$, 
and can be evaluated numerically
by using the transformation of the Jacobi coordinates \cite{bb}.
The completeness of the CS basis guarantees the convergence of the method
with increasing $N$ and angular momentum channels.

Now, by applying the bra $\langle \widetilde{ n'' \nu'' l'' \lambda''}|$
on Eq.\ (\ref{fmpa}) from left, the solution of the inhomogeneous
Faddeev-Merkuriev equation
turns into the solution of a matrix equation for the component vector
$\underline{\psi}=
 \mbox{}_1 \langle \widetilde{ n\nu l\lambda} | \psi  \rangle$
\begin{equation}
 \underline{\psi} =  \underline{\Phi}_1^{(l)} + \underline{G}_1^{(l)}  
\underline{v}_1^{(s)}   \underline{\psi} , \label{fn-eq1sm}  
\end{equation}
where 
\begin{equation}
\underline{\Phi}_1^{(l)} = \mbox{}_1 \langle \widetilde{ n\nu l\lambda } 
|\Phi_1^{(l)} \rangle
\end{equation}
and
\begin{equation}
\underline{G}_1^{(l)}=\mbox{}_1 \langle \widetilde{
n\nu l\lambda} |G_1^{(l)}|\widetilde{n' \nu' l' \lambda'}\rangle_1.
\end{equation}
The formal solution of Eq.\ (\ref{fn-eq1sm}) is given by
\begin{equation}  \label{fep1}
\underline{\psi }= \lbrack (\underline{G}_1^{(l)})^{-1}-
\underline{v}_1^{(s)}\rbrack^{-1} (\underline{G}_1^{(l)})^{-1}  
\underline{\Phi}_1^{(l)}.
\end{equation}

Unfortunately neither   $\underline{G}_1^{(l)}$ nor  $\underline{\Phi}_1^{(l)}$ 
are known. They are related to the 
Hamiltonian $H_1^{(l)}$, which is still a complicated three-body Coulomb
Hamiltonian. The approximation scheme for 
$\underline{G}_1^{(l)}$ and $\underline{\Phi}_1^{(l)}$
is presented  in Ref.\ \cite{phhky}.
Starting from the resolvent relation
\begin{equation}
G_1^{(l)}=\widetilde{G}_1 + \widetilde{G}_1 U_1 G_1^{(l)},
\label{ls1}
\end{equation}
where $\widetilde{G}_1$ is the resolvent operator of the Hamiltonian
\begin{equation} \label{htilde}
\widetilde{H}_1 = H^{0}+v_1^C
\end{equation}
and the potential $U_1$ is defined by
\begin{equation}
U_1=v_2^{(l)}+v_3^C,
\end{equation}
for the CS matrix elements $(\underline{G}^{(l)}_1)^{-1}$ we get
\begin{equation}
(\underline{G}^{(l)}_1)^{-1}= 
(\underline{\widetilde{G}}_1)^{-1} - \underline{U}_1,
\label{gleq}
\end{equation}
where 
\begin{equation}
\underline{\widetilde{G}}_{1} =
 \mbox{}_1\langle \widetilde{n \nu l \lambda} | 
 \widetilde{G}_1 | \widetilde{ n' \nu' l' \lambda'} \rangle_1  
 \label{gtilde}
\end{equation}
and 
\begin{equation}
\underline{U}_{1} =
 \mbox{}_1\langle n\nu l \lambda | U_1 | n' \nu' l' \lambda' \rangle_1.
\end{equation}
These  latter matrix elements can again be evaluated numerically.

Similarly, also the wave function $|{\Phi}_1^{(l)}\rangle$,
a scattering eigenstate of $H_1^{(l)}$, satisfies the
Lippmann-Schwinger equation
\begin{equation}
|{\Phi}_1^{(l)}\rangle=|\widetilde{\Phi}_1\rangle + 
\widetilde{G}_1 U_1   |{\Phi}_1^{(l)}\rangle,
\label{eqlsphil}
\end{equation}
where  $|\widetilde{\Phi}_1 \rangle$ is an eigenstate of 
$\widetilde{H}_1$. The solution is given by
\begin{equation}
\underline{\Phi}_1^{(l)} = 
[(\underline{\widetilde{G}}_1)^{-1} - \underline{U}^1]^{-1}
(\underline{\widetilde{G}}_1)^{-1} \underline{\widetilde{\Phi}}_1,
\label{eqphil}
\end{equation}
where $\underline{\widetilde{\Phi}}_{1}  = 
\mbox{}_1\langle \widetilde{ n \nu l \lambda }| 
\widetilde{\Phi}_1 \rangle$.

The three-particle free Hamiltonian
can be written  as a sum of two-particle free Hamiltonians 
\begin{equation}
H^0=h_{x_1}^0+h_{y_1}^0.
\end{equation}
Consequently the Hamiltonian $\widetilde{H}_1$ of Eq.\ (\ref{htilde}) 
appears as a sum of two two-body Hamiltonians acting on different coordinates 
\begin{equation}
\widetilde{H}_1 =h_{x_1}+h_{y_1},
\end{equation}
with $h_{x_1}=
h_{x_1}^0+v_1^C(x _1)$ and $h_{y_1}=h_{y_1}^0$, which, of course, commute.
Therefore the eigenstates of $\widetilde{H}_1$, in CS representation, 
are given by
\begin{equation}
\mbox{}_1\langle \widetilde{ n \nu l \lambda }| 
\widetilde{\Phi}_1 \rangle = \mbox{}_1\langle \widetilde{ n l}| 
{\phi}_1 \rangle \times \mbox{}_1\langle \widetilde{ \nu \lambda }| 
{\chi}_1 \rangle,
\label{phichi}
\end{equation}
where $|\phi_1 \rangle$ and $|\chi_1 \rangle$ are bound and scattering 
eigenstates of $h_{x_1}$ and $h_{y_1}$, respectively.
The CS matrix elements of the two-body  bound and scattering states 
$\langle \widetilde{ n l}| 
{\phi} \rangle$ and $\langle \widetilde{ \nu \lambda }| 
{\chi} \rangle$, respectively, 
are know analytically from the two-body case \cite{cspse2}.
 
The most crucial point in this procedure is the
 calculation of the matrix elements
$\underline{\widetilde{G}}_1$.
The Green's operator $\widetilde{G}_1$
is a resolvent of the sum of two commuting Hamiltonians.
Thus, according to the convolution theorem, the three-body Green's operator
$\widetilde{G}_1$ equates to
a convolution integral of two-body Green's operators, i.e.
\begin{equation}
\widetilde{G}_1 (z)=
\frac 1{2\pi {i}}\oint_C  dz' \,g_{x_1}(z-z')\;g_{y_1}(z'),
\label{contourint}
\end{equation}
where
$g_{x_1}(z)=(z-h_{x_1})^{-1}$  and
$g_{y_1}(z)=(z-h_{y_1})^{-1}$.
The contour $C$ should be taken  counterclockwise
around the singularities of $g_{y_1}$
such a way that $g_{x_1}$ is analytic on the domain encircled
by $C$.

\begin{figure}
\includegraphics[width=0.45\textwidth]{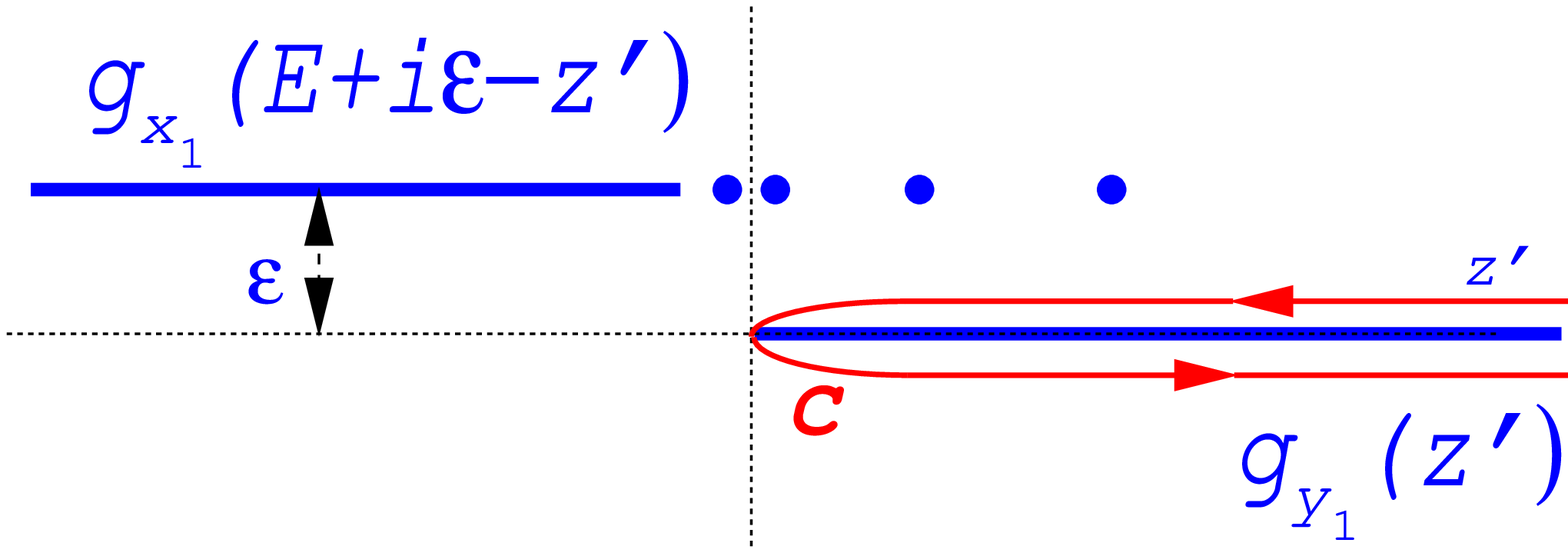}
\caption{Analytic structure of $g_{x_1}(E+{\mathrm{i}}\varepsilon-z')\;
g_{y_1}(z')$ as a function of $z'$, $\varepsilon>0$. The Green's operator
$g_{y_1}(z')$ has a branch-cut on the $[0,\infty)$ interval, while
$g_{x_1}(E+{\mathrm{i}}\varepsilon-z')$ has a branch-cut on the 
$(-\infty,E+{\mathrm{i}}\varepsilon]$ interval and infinitely many 
poles accumulated at $E+{\mathrm{i}}\varepsilon$ (denoted by dots).
The contour $C$ encircles the branch-cut of
$g_{y_1}$. In the $\varepsilon \to 0$ limit the singularities
of $g_{x_1}(E+{\mathrm{i}}\varepsilon -z')$ would penetrate
into the area covered by $C$. }
\label{fig1}
\end{figure}

\begin{figure}
\includegraphics[width=0.45\textwidth]{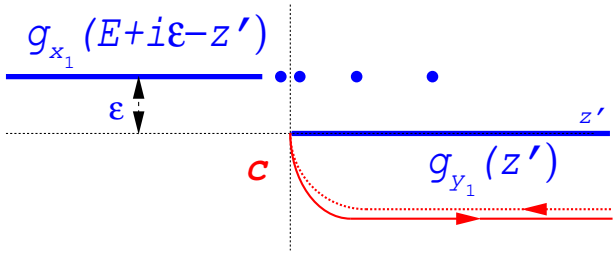}
\caption{The contour of Fig.\ \ref{fig1} is
deformed analytically such that a part of it goes on the unphysical
Riemann-sheet of $g_{y_1}$ (drawn by broken line) and the other part detoured
away from the cut. Now, the contour avoids the singularities of 
$g_{x_1}(E+{\mathrm{i}}\varepsilon-z')$ even in the
 $\varepsilon \to 0$ limit.}
\label{fig2}
\end{figure}

In the time-independent scattering theory the Green's operator has a branch-cut
singularity at scattering energies. In our formalism $\widetilde{G}_1 (E)$
should be understand as 
$\widetilde{G}_1 (E)=\lim_{\varepsilon\to 0} \widetilde{G}_1 
(E +{\mathrm{i}}\varepsilon)$, with $\varepsilon > 0$, and $E < 0$, since
in this work we are considering scattering below the three-body breakup 
threshold.
To examine the analytic structure of the integrand in Eq.\ (\ref{contourint}) 
let us take $\varepsilon$ finite. By doing so,
the singularities of $g_{x_1}$ and $g_{y_1}$ become well separated.
In fact, $g_{y_1}$ is a free Green's operator with branch-cut on 
the $[0,\infty)$ interval, while
$g_{x_1}(E+{\mathrm{i}}\varepsilon-z')$ is a Coulomb Green's operator, which,
as function of $z'$, has a branch-cut on the 
$(-\infty,E+{\mathrm{i}}\varepsilon]$ interval and  infinitely many 
poles accumulated at $E+{\mathrm{i}}\varepsilon$.
Now, the branch cut of $g_{y_1}$ can easily be encircled such  that
the singularities of $g_{x_1}$ lie outside the encircled domain 
(Fig.~\ref{fig1}). However, this would not be the case 
in the $\varepsilon\to 0$ limit.
Therefore the contour $C$ is deformed analytically
such that the upper part descends into the unphysical
Riemann sheet of $g_{y_1}$, while the lower part of $C$ is detoured away 
from the cut (Fig.~\ref{fig2}). The contour in Fig.~\ref{fig2} 
is achieved by deforming 
analytically the one in Fig.~\ref{fig1}, but now, even in 
the  $\varepsilon\to 0$ limit,
the contour in Fig.~\ref{fig2} avoids the singularities of $g_{x_1}$.
Thus, with the contour in Fig.~\ref{fig2} the mathematical conditions for
the contour integral representation of $\widetilde{G}_1$ in
Eq.~(\ref{contourint}) is met also for scattering-state energies. 
The matrix elements $\underline{\widetilde{G}}_1$
can be cast into the form
\begin{equation}
\widetilde{\underline{G}}_1 (E )= \frac 1{2\pi \mathrm{i}}\oint_C
dz' \,\underline{g}_{x_1 }(E -z')\; \underline{g}_{y_1}(z'),
\label{contourint2}
\end{equation}
where the corresponding CS matrix elements of the two-body Green's operators in
the integrand are known analytically for all complex energies 
\cite{cspse2,phhky}.

In the three-potential formalism \cite{pzsc,phhky} the $S$ matrix can be 
decomposed into three terms. The first one describes a single channel Coulomb 
scattering, the second one is a multichannel two-body-type scattering due
to the potential $U$, and  the third one is a genuine three-body
scattering. In our  $e^- + H$ case the target is neutral and the first term 
is absent. For the on-shell $T$ matrix we have
\begin{equation}
T_{f i} =  \sqrt{\frac{\mu_f \mu_i}{k_f k_i}} \left(\langle 
\widetilde{\Phi }_{1 f}^{(-)}|U_1 |\Phi _{1 i}^{(l)(+)}\rangle +
\langle \Phi _{1 f}^{(l)(-)}|v_1^{(s)} | \psi _{2 i}^{(+)}\rangle\right),
\label{s3}
\end{equation}
where $i$ and $f$ refer to the initial and the final states, respectively, 
$\mu$ is the channel reduced mass and $k$ is the channel wave number.
Having the solutions $\underline{\psi}$ and $\underline{\Phi}^{(l)}$ and the
matrix elements $\underline{U}_1$ and $\underline{v}^{(s)}_1$, the
$T$ matrix elements can easily be evaluated. The spin-weighted 
cross section of the transition $i\to f$ is given by
\begin{equation}
\sigma_{f i} = \frac{\pi a_0^2}{k_i^2} \frac{(2 S_{12}+1)(2L+1)}
{(2 l_i +1)} |T_{f i}|^2,
\end{equation}
where $a_0$ is the Bohr radius, $L$ is the total angular momentum,
$S_{12}$ is the total spin of the two electrons  and $l_i$ is the 
angular momentum of the target hydrogen atom.

\section{Results}

In the numerical calculations we use atomic units 
(the mass of the electrons $m_1=m_2=1$ and 
the mass of the proton $m_3=1836.151527$). 
In this paper we are concerned
with total angular momenta $L=0$ and $L=1$.
The formula (\ref{s3}) gives some hint for the choice of the parameters
in the splitting function $\zeta$. We can expect good convergence if the
''size'' of $v_1^{(s)}$ corresponds to the ''size'' of $\Phi _{1 f}^{(l)(-)}$.
Therefore we may need to adjust the parameters of the splitting function
if we consider more and more open channels. Consequently, we also need to
adjust the $b$ parameter of the CS basis. We found that the final results and
the rate of the convergence does not depend on the choice of $b$, within
a rather broad interval around the optimal value.

Having the $T$ matrix we can also calculate the 
$K$ matrix, whose symmetry, which is equivalent to the unitarity of the $S$
matrix, provides a delicate and independent test of the method. 
We observed that if either the parameters of the splitting function 
are too far from the optimum or the
convergence with the basis is not achieved the $K$ matrix fails to be symmetric.
In the separable expansion we take up to $9$ bipolar angular momentum channels
with CS functions up to $N=36$. This requires solution of
complex general matrix equations with maximal size of $12321 \times 12321$, 
a problem which can even be handled on a workstation. We need relatively
small basis because in this approach we approximate only short-range type
potentials and the correct asymptotic is guaranteed by the Green's operators.

We present first our $S$-wave results for energies below the $H(n=2)$ threshold. 
In this energy region we use parameters $\nu=2.1$,
$x_0=3$, $y_0=20$ and $b=0.6$.
Table \ref{tab1} shows elastic phase shifts at several values of electron
momenta $k_1$. Our results, which was achieved by using finite proton mass,
 agree very well with variational calculations of 
Ref.\ \cite{schwarz}, $R$-matrix calculations of Ref.\ \cite{scholz},
finite-element method of Ref.\ \cite{botero},
as well as with the results of direct numerical solution
of the Schr\"odinger equation of Ref.\ \cite{wang}, 
where infinite mass for proton were adopted. 
We also compare our calculation with the differential equation solution 
of the modified Faddeev equations \cite{kwh}. We can observe perfect agreements
with all the previous calculations.

In Table \ref{tab2} we present $S$-wave partial cross sections and $K$ matrices
between the $H(n=2)-H(n=3)$ thresholds at channel energy  $E_1=0.81\text{Ry}$ 
and for $L=0$, where we have $3$ open channels. 
We used parameters $\nu=2.1$, $x_0=3.5$, 
$y_0=20$ and $b=0.3$. For comparison we also show the results of a
configuration-space Faddeev calculation \cite{hu}. We can report perfect
agreements not only for the cross sections but also for the $K$ matrix
(except for an unphysical phase factor). Our cross sections are also in a good
agreements with the results of Ref.\ \cite{wang}.

In Tables \ref{tab3} we show the $S$-wave $K$ matrix
between the $H(n=3)-H(n=4)$  thresholds at channel energy $E_1=0.93\text{Ry}$, 
where we have $6$ open channels. We used parameters $\nu=2.1$, $x_0=4$, 
$y_0=20$ and $b=0.2$. We can see that the $K$ matrix is nearly perfectly
symmetric. In Tables \ref{tab4} we present $S$-wave  partial cross sections 
between the $H(n=3)-H(n=4)$  thresholds at channel energies $E_1=0.93\text{Ry}$, 
$E_1=0.91\text{Ry}$ and $E_1=0.89\text{Ry}$, respectively. 
In Tables \ref{tab5}-\ref{tab8} we present the corresponding $P$-wave $K$ matrices
and cross sections.

\section{Summary}

In this work we have studied electron-hydrogen scattering problem
by solving the Faddeev-Merkuriev integral equations. In this particular case,
where two particles are identical, the Faddeev scheme results in an
one-component equation, which, however, gives full account on the asymptotic 
and symmetry properties of the system. We solved the integral
equations by applying the Coulomb-Sturmian separable expansion method.

We calculated $S$- and $P$-wave 
scattering and reaction cross sections for energies up to the
$H(n=4)$ threshold. Our nearly perfectly symmetric $K$ matrices indicate that 
in our approach all the fine details of the scattering processes are properly 
taken into account.

\begin{acknowledgments}
This work has been supported by the NSF Grant No.Phy-0088936
and by the OTKA Grants No.\ T026233 and No.\ T029003. 
We are thankful to the Aerospace Engineering Department of CSULB
for the generous allocation of computer resources.
\end{acknowledgments}

\begin{table}[!htb]
\caption{Singlet ($\mbox{}^1S^e,p=+1$) and triplet ($\mbox{}^3S^e,p=-1$)
phase shifts of elastic $S$-wave $e^- + H$ scattering .}
\label{tab1}
\begin{tabular}{lcccccc}
$k$&  Ref.\ \cite{schwarz} &  Ref.\ \cite{scholz} &  Ref.\ \cite{botero} & 
Ref.\ \cite{wang} & 
Ref.\ \cite{kwh} & This work \\ \hline
\multicolumn{6}{c}{ $\mbox{}^1S^e, p=+1$ } \\ \hline 
0.1 & 2.553 & 2.550 & 2.553 & 2.555 & 2.553 & 2.552 \\
0.2 & 2.0673& 2.062 & 2.066 & 2.066 & 2.065 & 2.064 \\
0.3 & 1.6964& 1.691 & 1.695 & 1.695 & 1.694 & 1.693 \\
0.4 & 1.4146& 1.410 & 1.414 & 1.415 & 1.415 & 1.412 \\
0.5 & 1.202 & 1.196 & 1.202 & 1.200 & 1.200 & 1.197  \\
0.6 & 1.041 & 1.035 & 1.040 & 1.041 & 1.040 & 1.037  \\
0.7 & 0.930 & 0.925 & 0.930 & 0.930 & 0.930 & 0.927 \\
0.8 & 0.886 &       & 0.887 & 0.887 & 0.885 & 0.884 \\ \hline
\multicolumn{6}{c}{ $\mbox{}^3S^e$, p=-1 } \\ \hline 
0.1 & 2.9388& 2.939 & 2.938 & 2.939 & 2.939 & 2.938 \\
0.2 & 2.7171& 2.717 & 2.717 & 2.717 & 2.717 & 2.717  \\
0.3 & 2.4996& 2.500 & 2.500 & 2.500 & 2.499 & 2.499  \\
0.4 & 2.2938& 2.294 & 2.294 & 2.294 & 2.294 & 2.294  \\
0.5 & 2.1046& 2.105 & 2.104 & 2.104 & 2.105 & 2.104 \\
0.6 & 1.9329& 1.933 & 1.933 & 1.933 & 1.933 & 1.932 \\
0.7 & 1.7797& 1.780 & 1.780 & 1.780 & 1.779 & 1.779 \\
0.8 & 1.643 &       & 1.645 & 1.644 & 1.641 & 1.643 \\ \hline 
\end{tabular}
\end{table}

\begin{table}[!htb]
\caption{$L=0$ partial cross sections (in $\pi a_0^2$) 
in the $H(n=2)-H(n=3)$ gap at channel energy $E_1=0.81 \text{Ry}$. 
Channel numbers $1$, $2$ and 
$3$ refer to the channels $e^- + H(1s)$, $e^- + H(2s)$ and 
$e^- + H(2p)$, respectively. For comparison the result of a 
configuration-space Faddeev calculation is presented.  }
\label{tab2}
\begin{tabular}{lcccc}
 & Ch.\# & 1 & 2  & 3   \\ \hline
\multicolumn{5}{c}{$\mbox{}^1S^e, p=+1 $} \\ \hline
\multicolumn{5}{c}{This work} \\ \hline
             & 1 & 0.564 & 0.061 & 0.024 \\
$\sigma_{ij}$ & 2 & 0.817 & 8.373 & 2.588 \\
             & 3 & 0.107 & 0.863 & 1.722 \\ \hline
             & 1 & 1.895 & -2.036 &  1.792 \\
$K_{ij}$     & 2 &-2.043 &  5.230 & -4.114 \\
             & 3 & 1.798 & -4.114 & 2.366 \\ \hline
\multicolumn{5}{c}{Method of Ref.\ \cite{hu}} \\ \hline
             & 1 & 0.568 & 0.061 & 0.024 \\
$\sigma_{ij}$ & 2 & 0.814 & 8.720 & 2.471 \\
             & 3 & 0.105 & 0.824 & 1.697 \\ \hline
             & 1 & 1.864 & 1.971 &  -1.671 \\
$K_{ij}$     & 2 & 1.980 &  5.131 & -3.843 \\
             & 3 & -1.679 & -3.843 & 2.028 \\ \hline
\multicolumn{5}{c}{$\mbox{}^3S^e, p=-1 $} \\ \hline
\multicolumn{5}{c}{This work} \\ \hline
               & 1 & 3.694 & 0.001  & 0.0006  \\ 
$\sigma_{ij}$  & 2 & 0.016 & 10.04 & 1.641  \\
               & 3 & 0.003 & 0.547 & 11.85  \\ \hline
               & 1 & 21.34 & 0.3255  & 0.6386  \\ 
$ K_{ij}$      & 2 & 0.3268 & -0.4404 & -0.4161  \\
               & 3 & 0.6409 & -0.4161 & 1.755  \\ \hline
\multicolumn{5}{c}{Method of Ref.\ \cite{hu}} \\ \hline
               & 1 & 3.696 & 0.001  & 0.0006  \\ 
$\sigma_{ij}$  & 2 & 0.016 & 10.20 & 1.678  \\
               & 3 & 0.003 & 0.560 & 11.77  \\ \hline
               & 1 & 24.76 & -0.3823  & -0.7510  \\ 
$ K_{ij}$      & 2 & -0.3803 & -0.4441 & -0.4167  \\
               & 3 & -0.7453 & -0.4165 & 1.737  \\ \hline
\end{tabular}
\end{table}

\begin{table}[!htb]
\caption{$S$-wave $K$ matrices
in the $H(n=3)-H(n=4)$ gap at channel energy $E_1=0.93\text{Ry}$.
The channel numbers $1$, $2$, $3$, $4$, $5$ and $6$ 
refer to the
channels $e^-(\lambda=0) + H(1s)$, $e^-(\lambda=0) + H(2s)$, 
$e^-(\lambda=0) + H(3s)$,
$e^-(\lambda=1) + H(1p)$, $e^-(\lambda=1) + H(2p)$ and 
$e^-(\lambda=2) + H(1d)$, respectively.}
\label{tab3}
\begin{tabular}{cllllll}
  Ch.\# & 1 & 2  & 3  & 4 & 5 & 6 \\ \hline
\multicolumn{7}{c}{$E_1=0.93\text{Ry}$, $\mbox{}^1S^e, p=+1 $} \\ \hline
 1 & 1.076 & -0.647 & -0.160 & 0.229 & 0.180 & 0.074  \\
 2 & -0.652 & 1.541 & -0.028 & 0.129 & 0.531 & 0.265 \\
 3 & -0.160 & -0.029 & 0.766 & 0.314 & -0.757 & -0.385 \\
 4 & 0.230 & 0.130 & 0.314 & -0.566 & -0.525 & -0.284 \\
 5 & 0.180 & 0.534 & -0.757  & -0.526 & 0.237 & 0.760 \\
 6 & 0.074 & 0.266 & -0.385 & -0.285 & 0.760 & 1.342 \\ \hline
\multicolumn{7}{c}{$E_1=0.93\text{Ry}$, $\mbox{}^3S^e, p=-1 $} \\ \hline
 1 & 9.054 &  0.507 &  0.019 &  0.666 &  0.099 &  0.028  \\
 2 & 0.543 & -1.700 & -0.111 & -1.530 & -0.113 & -0.120 \\
 3 & 0.025 & -0.112 &  0.155 & -0.050 & -0.926 & -0.070 \\
 4 & 0.702 & -1.532 & -0.050 & -0.851 & -0.253 & -0.048 \\
 5 & 0.104 & -0.114 & -0.926 & -0.253 &  0.927 &  0.449 \\
 6 & 0.030 & -0.120 & -0.070 & -0.049 &  0.449 & -0.111 \\ \hline
\end{tabular}
\end{table}

\begin{table}[!htb]
\caption{$L=0$ partial cross sections (in $\pi a_0^2$) 
in the $H(n=3)-H(n=4)$ gap at channel energies $E_1=0.93\text{Ry}$, 
$E_1=0.91\text{Ry}$ and $E_1=0.89\text{Ry}$, respectively. 
The channel numbers $1$, $2$, $3$, $4$, $5$ and $6$ 
refer to the
channels $e^-(\lambda=0) + H(1s)$, $e^-(\lambda=0) + H(2s)$, 
$e^-(\lambda=0) + H(3s)$,
$e^-(\lambda=1) + H(1p)$, $e^-(\lambda=1) + H(2p)$ and 
$e^-(\lambda=2) + H(1d)$, respectively.}
\label{tab4}
\begin{tabular}{cllllll}
  Ch.\# & 1 & 2  & 3  & 4 & 5 & 6 \\ \hline
\multicolumn{7}{c}{$E_1=0.93\text{Ry}$, $\mbox{}^1S^e, p=+1 $} \\ \hline
 1 &0.44 & 0.48(-1) & 0.67(-2) & 0.28(-1) & 0.86(-2) & 0.20(-2)  \\
 2 &0.25 & 3.02 & 0.19(-1) & 0.10 & 0.12 & 0.40(-1)\\
 3 &0.15 & 0.83(-1) & 4.68 & 0.71 & 2.41 & 0.86 \\
 4 &0.49(-1) & 0.34(-1) & 0.55(-1) & 0.49 & 0.59(-1) & 0.24(-1) \\
 5 &0.65(-1) & 0.18 & 0.80  & 0.26 & 1.48 & 0.44 \\
 6 &0.89(-2) & 0.35(-1) & 0.17 & 0.61(-1) & 0.27 & 2.0\\ \hline
\multicolumn{7}{c}{$E_1=0.93\text{Ry}$, $\mbox{}^3S^e, p=-1 $} \\ \hline
 1 & 3.18 & 0.22(-2) & 0.43(-4) & 0.21(-2) & 0.26(-4) & 0.14(-5) \\
 2 & 0.12(-1) & 5.92 & 0.93(-2) & 3.77 & 0.44(-1) & 0.61(-1) \\
 3 & 0.97(-3) & 0.40(-1) & 7.56 & 0.35 & 11.6 & 3.34 \\
 4 & 0.39(-2) & 1.26 &  0.26(-1) & 0.87 & 0.11(-1) & 0.19(-2) \\
 5 & 0.23(-3) & 0.63(-1) & 3.87 & 0.48(-1) & 9.14 & 1.07 \\
 6 & 0.79(-5) & 0.53(-1) & 0.67 & 0.49(-2) & 0.64 & 0.34 \\  \hline
\multicolumn{7}{c}{$E_1=0.91\text{Ry}$, $\mbox{}^1S^e, p=+1 $} \\ \hline
 1 &0.46 & 0.45(-1) & 0.90(-2) & 0.24(-1) & 0.89(-2) & 0.18(-2)  \\
 2 &0.26 & 3.74 & 0.24 & 0.77(-1) & 0.74(-1) & 0.15(-1)\\
 3 &0.38 & 1.77 & 5.46 & 1.11 & 0.90 & 1.14 \\
 4 &0.46(-1) & 0.26(-1) & 0.50(-1) & 0.49 & 0.86(-1) & 0.30(-1) \\
 5 &0.13 & 0.18 & 0.30  & 0.64 & 11.5 & 0.37 \\
 6 &0.16(-1) & 0.23(-1) & 0.23 & 0.13 & 0.22 & 1.15 \\ \hline
\multicolumn{7}{c}{$E_1=0.91\text{Ry}$, $\mbox{}^3S^e, p=-1 $} \\ \hline
 1 & 3.26  & 0.22(-2) & 0.23(-4) & 0.20(-2) & 0.20(-4) & 0.17(-5) \\
 2 & 0.13(-1) & 7.22 & 0.24(-1) & 4.07 & 0.28(-1) & 0.25(-1) \\
 3 & 0.96(-3) & 0.18 & 9.11 & 0.33 & 0.27 & 12.25 \\
 4 & 0.37(-2) & 1.36 &  0.15(-1) & 0.98 & 0.10(-1) & 0.31(-2) \\
 5 & 0.26(-3) & 0.69(-1) & 0.89(-1) & 0.74(-1) & 44.97 & 0.44 \\
 6 & 0.14(-4) & 0.38(-1) & 2.45 & 0.14(-1) & 0.27 & 2.81 \\  \hline
\multicolumn{7}{c}{$E_1=0.89\text{Ry}$, $\mbox{}^1S^e, p=+1 $} \\ \hline
 1 &0.48 & 0.47(-1) & 0.53(-2) & 0.21(-1) & 0.79(-2) & 0.26(-2)  \\
 2 &0.30 & 4.67 & 0.32(-1) & 0.61(-1) & 0.14 & 0.12 \\
 3 &2.98 & 2.80 & 259.8 & 19.02 & 1.58 & 7.12 \\
 4 &0.45(-1) & 0.20(-1) & 0.72(-1) & 0.51 & 0.77(-1) & 0.13(-1) \\
 5 &1.48 & 3.40 & 0.53  & 6.78 & 119.8 & 2.67 \\
 6 &0.29 & 2.04 & 1.42 & 0.70 & 1.60 & 38.90 \\ \hline
\multicolumn{7}{c}{$E_1=0.91\text{Ry}$, $\mbox{}^3S^e, p=-1 $} \\ \hline
 1 & 3.34  & 0.22(-2) & 0.67(-5) & 0.17(-2) & 0.90(-5) & 0.25(-5) \\
 2 & 0.13(-1) & 8.68 & 0.94(-2) & 4.33 & 0.17(-1) & 0.87(-1) \\
 3 & 0.37(-3) & 0.83 & 1321.0 & 1.75 & 152.8 & 58.26 \\
 4 & 0.33(-2) & 1.44 &  0.66(-2) & 1.25 & 0.67(-2) & 0.12(-2) \\
 5 & 0.16(-2) & 0.49 & 50.93 & 0.59 & 124.6 & 56.47 \\
 6 & 0.25(-3) & 0.15 & 11.65 & 0.63(-1) & 33.88 & 218.4 \\  \hline
\end{tabular}
\end{table}

\widetext

\begin{table}[!htb]
\caption{$P$-wave $K$ matrices
in the $H(n=3)-H(n=4)$ gap at channel energy $E_1=0.93\text{Ry}$.
The channel numbers $1$, $2$, $3$, $4$, $5$, $6$, $7$, $8$ and $9$
refer to the channels 
$e^-(\lambda=1) + H(1s)$, $e^-(\lambda=1) + H(2s)$, $e^-(\lambda=1) + H(3s)$,
$e^-(\lambda=0) + H(2p)$, $e^-(\lambda=0) + H(3p)$,
$e^-(\lambda=2) + H(2p)$, $e^-(\lambda=2) + H(3p)$,
$e^-(\lambda=1) + H(3d)$ and $e^-(\lambda=3) + H(3d)$, respectively.}
\label{tab5}
\begin{tabular}{clllllllll}
  Ch.\# & 1 & 2  & 3  & 4 & 5 & 6 & 7 & 8 & 9 \\ \hline
\multicolumn{7}{c}{$E_1=0.93\text{Ry}$, $\mbox{}^1S^e, p=+1 $} \\ \hline
 1 & -1.888 & -7.518 & 13.24 & 9.699 & 8.320 & 7.148 & 1.992 & 4.684 & 30.96 \\
 2 & -7.525 & -29.70 & 51.80 & 38.14 & 32.73 & 28.88 & 7.839 & 18.28 & 121.5 \\
 3 &  13.30 & 51.99 &-89.98 & -67.93 & -56.77 & -50.01 & -13.90 & -29.57 & -216.6 \\
 4 & 9.665 & 37.98 & -67.40 & -48.21 & -42.35 & -36.39 & -9.947 & -24.07 & -156.7 \\
 5 & 8.346 & 32.81 & -56.70 & -42.64 & -36.30 & -31.16 & -9.349 & -19.40 & -136.0 \\
 6 & 7.151 & 28.87 & -49.82 & -36.54 & -31.08 & -28.11 & -7.718 & -17.41 & -117.3 \\
 7 & 2.006 & 7.885 & -13.94 & -10.05 & -9.381 & -7.765 & -2.651 & -4.874 & -34.08 \\
 8 & 4.755 & 18.64 & -29.92 & -24.51 & -19.64 & -17.67 & -4.915 & -8.953 & -73.21 \\
 9 & 31.01 & 121.6 & -215.9 & -157.5 & -135.7 & -117.3 & -33.89 & -72.15 & -510.6 \\
 \hline
\multicolumn{7}{c}{$E_1=0.93\text{Ry}$, $\mbox{}^3S^e, p=-1 $} \\ \hline
 1 &  0.454 & -0.303 & -0.051 & -0.020 &  0.080 &  0.043 & -0.017 &  0.149 &  0.128 \\ 
 2 & -0.301 & -2.453 & -0.669 &  0.383 &  0.552 &  1.112 &  0.017 &  1.145 &  1.060 \\
 3 & -0.051 & -0.672 &  0.398 & -0.465 &  1.140 & -0.371 &  0.0001 & 0.578 &  0.486 \\
 4 & -0.020 &  0.382 & -0.464 &  0.354 & -1.133 & -0.236 &  0.883 & -0.528 & -0.110 \\
 5 &  0.079 &  0.553 &  1.137 & -1.136 &  3.936 & -0.699 & -3.202 &  1.075 & -0.989 \\ 
 6 &  0.041 &  1.113 & -0.372 & -0.236 & -0.701 &  0.289 &  0.520 & -0.769 & -0.456 \\ 
 7 & -0.016 &  0.018 &  0.002 &  0.884 & -3.203 &  0.518 &  1.673 & -1.484 & -0.226 \\
 8 &  0.148 &  1.147 &  0.576 & -0.530 &  1.075 & -0.769 & -1.483 & -0.055 & -0.278 \\
 9 &  0.127 &  1.062 &  0.486 & -0.111 & -0.988 & -0.457 & -0.226 & -0.277 &  0.090 \\
\hline
\end{tabular}
\end{table}

\begin{table}[!htb]
\caption{$L=1$ partial cross sections (in $\pi a_0^2$ unit)
 in the $H(n=3)-H(n=4)$ gap at channel energy $E_1=0.93\text{Ry}$.
The channel numbers $1$, $2$, $3$, $4$, $5$, $6$, $7$, $8$ and $9$
refer to the channels 
$e^-(\lambda=1) + H(1s)$, $e^-(\lambda=1) + H(2s)$, $e^-(\lambda=1) + H(3s)$,
$e^-(\lambda=0) + H(2p)$, $e^-(\lambda=0) + H(3p)$,
$e^-(\lambda=2) + H(2p)$, $e^-(\lambda=2) + H(3p)$,
$e^-(\lambda=1) + H(3d)$ and $e^-(\lambda=3) + H(3d)$, respectively. }
\label{tab6}
\begin{tabular}{clllllllll}
  Ch.\# & 1 & 2  & 3  & 4 & 5 & 6 & 7 & 8 & 9 \\ \hline
\multicolumn{7}{c}{$E_1=0.93\text{Ry}$, $\mbox{}^1S^e, p=+1 $} \\ \hline
 1 & 0.380(-2) & 0.104(-1) &  0.138(-2) & 0.394(-1) &  0.677(-2) & 0.125(-1) &  0.543(-2) &
 0.664(-2) & 0.180(-2) \\
 2 & 0.530(-1) &  0.208(1) &  0.760(-2) & 0.152(1) &  0.139  &  0.135(1) &  0.284(-1) &  0.450  & 
 0.103  \\ 
 3 & 0.319(-1) &  0.321(-1) &  0.311(2) &  0.117(1) &  0.340(1) &  0.170  &  0.459(1) &  0.191(1)
 &  0.282(1) \\
 4 & 0.679(-1) &  0.506  &  0.903(-1) &  0.157(1) &  0.104 &  0.151  &  0.213  &  0.169  & 
 0.796(-1) \\ 
 5 & 0.508(-1) &  0.201  &  0.113(1) &  0.450  &  0.415(1) &  0.103(1) &  0.169(1) &  0.113(1) & 
 0.871(-1) \\ 
 6 & 0.219(-1) &  0.448  &  0.131(-1) &  0.150  &  0.235  &  0.164(1) &  0.647(-1) &  0.399(-1) & 
 0.183(-2) \\
 7 & 0.412(-1) &  0.415(-1) &  0.153(1) &  0.928  &  0.169(1) &  0.282  &  0.335(1) &  0.105  & 
 0.233  \\ 
 8 & 0.296(-1) &  0.391  &  0.383  &  0.440  &  0.679  &  0.105  &  0.620(-1) &  0.800(1) & 
 0.283  \\ 
 9 & 0.807(-2) & 0.890(-1) &  0.562  &  0.208  &  0.523(-1) &  0.468(-2) & 0.139  &  0.283  &  0.116(2) \\
\multicolumn{7}{c}{$E_1=0.93\text{Ry}$, $\mbox{}^3S^e, p=-1 $} \\ \hline
 1 &  0.178(1) & 0.484(-1) & 0.853(-2) & 0.158(-1) & 0.603(-2) & 0.167(-1) & 0.457(-2) & 0.191(-2) & 0.625(-3) 
\\
 2 &  0.247 &  0.235(2) &  0.390 &  0.109(1) & 0.435 &  0.294(1) & 0.163(1) & 0.171(1) & 0.182(1) 
\\
 3 &  0.193 &  0.170(1) & 0.514(2) &  0.892(1) & 0.208(1) & 0.105(2) &  0.167(2) &  0.596 &  0.381(1) 
\\
 4 &  0.277(-1) & 0.362 &  0.683 &  0.846 &  0.576 &  0.801 &  0.344(-1) & 0.924(-2) & 0.920(-1) 
\\
 5 &  0.453(-1) & 0.633 &  0.695 &  0.251(1) & 0.373(2) &  0.525 &  0.348(1) & 0.295(1) & 0.367(1) 
\\
 6 &  0.291(-1) & 0.981 &  0.810 &  0.804 &  0.121 &  0.416(1) & 0.887(-1) & 0.121 &  0.803(-1) 
\\
 7 &  0.333(-1) & 0.236(1) & 0.556(1) & 0.151 &  0.348(1) & 0.388 &  0.276(2) &  0.290(1) & 0.186(1) 
\\
 8 &  0.831(-2) & 0.149(1) & 0.119 &  0.240(-1) & 0.177(1) & 0.315 &  0.174(1) & 0.448(1) & 0.280(1) 
\\
 9 &  0.259(-2) & 0.158(1) & 0.760 &  0.241 &  0.220(1) & 0.208 &  0.111(1) & 0.280(1) & 0.935(1) 
\\ \hline
\end{tabular}
\end{table}

\begin{table}[!htb]
\caption{The same as in Table \ref{tab6} at channel energy $E_1=0.91\text{Ry}$. }
\label{tab7}
\begin{tabular}{clllllllll}
  Ch.\# & 1 & 2  & 3  & 4 & 5 & 6 & 7 & 8 & 9 \\ \hline
\multicolumn{7}{c}{$E_1=0.91\text{Ry}$, $\mbox{}^1S^e, p=+1 $} \\ \hline
 1 &  0.365(-2) & 0.102(-1) & 0.102(-2) & 0.437(-1) & 0.474(-2) & 0.137(-1) & 0.397(-2) & 0.521(-2) & 0.154(-2) 
\\
 2 &  0.576(-1) & 0.218(1) & 0.648(-2) & 0.185(1) & 0.395 &  0.122(1) & 0.694(-1) & 0.109 &  0.276(-1) 
\\
 3 &  0.428(-1) & 0.485(-1) & 0.963(2) &  0.213(1) & 0.380(1) & 0.262 &  0.107(1) & 0.683(1) & 0.517(1) 
\\
 4 &  0.829(-1) & 0.617 &  0.954(-1) & 0.193(1) & 0.155 &  0.245 &  0.206 &  0.989(-1) & 0.279(-1) 
\\
 5 &  0.666(-1) & 0.981 &  0.127(1) & 0.115(1) & 0.509(1) & 0.605 &  0.326(1) & 0.168(1) & 0.944  
\\
 6 &  0.261(-1) & 0.408 &  0.118(-1) & 0.246 &  0.818(-1) & 0.188(1) & 0.367(-1) & 0.139 &  0.520(-1) 
\\
 7 &  0.563(-1) & 0.174 &  0.356 &  0.153(1) & 0.326(1) & 0.271 &  0.570(1) & 0.157(1) & 0.262(1) 
\\
 8 &  0.444(-1) & 0.163 &  0.137(1) & 0.443 &  0.101(1) & 0.616 &  0.940 &  0.160(2) &  0.438 
\\
 9 &  0.133(-1) & 0.411(-1) & 0.103(1) & 0.125 &  0.567 &  0.232 &  0.157(1) & 0.438 &  0.187(2) 
\\ \hline
\multicolumn{7}{c}{$E_1=0.91\text{Ry}$, $\mbox{}^3S^e, p=-1 $} \\ \hline
 1 &  0.182(1) & 0.438(-1) & 0.788(-2) & 0.1567(-1) & 0.605(-2) & 0.159(-1) & 0.450(-2) & 0.209(-2) & 0.633(-3)  
\\
 2 &  0.243 &  0.241(2) &  0.209(1) & 0.170(1) & 0.126(1) & 0.479(1) & 0.106(1) & 0.632 &  0.652  
\\
 3 &  0.329 &  0.156(2) &  0.166(3) &  0.888(1) & 0.146(1) & 0.851(1) & 0.334(1) & 0.471(1) & 0.797(1) 
\\
 4 &  0.296(-1) & 0.567 &  0.397 &  0.139(1) & 0.464 &  0.102(1) & 0.119 &  0.123 &  0.107 
\\
 5 &  0.846(-1) & 0.311(1) & 0.486 &  0.345(1) & 0.814(2) &  0.524 &  0.604(1) & 0.480 &  0.144(1) 
\\
 6 &  0.296(-1) & 0.160(1) & 0.382 &  0.102(1) & 0.706(-1) & 0.445(1) & 0.192 &  0.177 &  0.140
\\
 7 &  0.632(-1) & 0.263(1) & 0.111(1) & 0.885 &  0.604(1) & 0.143(1) & 0.540(2) &  0.255 &  0.113(1) 
\\
 8 &  0.179(-1) & 0.943 &  0.942 &  0.550 &  0.2895 &  0.793 &  0.153 &  0.633(1) & 0.231(1) 
\\
 9 &  0.551(-2) & 0.971 &  0.159(1) & 0.480 &  0.8620 &  0.627 &  0.676 &  0.231(1) & 0.218(2) 
\\ \hline
\end{tabular}
\end{table}

\begin{table}[!htb]
\caption{The same as in Table \ref{tab6} at channel energy $E_1=0.89\text{Ry}$. }
\label{tab8}
\begin{tabular}{clllllllll}
  Ch.\# & 1 & 2  & 3  & 4 & 5 & 6 & 7 & 8 & 9 \\ \hline 
\multicolumn{7}{c}{$E_1=0.89\text{Ry}$, $\mbox{}^1S^e, p=+1 $} \\ \hline
   1 &  0.342(-2) & 0.940(-2) & 0.819(-3) &  0.474(-1) & 0.262(-2) & 0.154(-1) & 0.264(-2) & 0.292(-2) & 0.849(-3)  
\\ 2 &  0.609(-1) & 0.252(1) & 0.412(-1) & 0.212(1) & 0.693(-1) & 0.108(1) & 0.976(-1) & 0.777(-1) & 0.306(-1) 
\\ 3 &  0.466 &  0.361(1) & 0.550(3) &  0.862(1) & 0.384(2) &  0.567(1) & 0.177(3) &  0.672(2) &  0.232(2) 
\\ 4 &  0.102 &  0.708 &  0.326(-1) & 0.233(1) & 0.126 &  0.422 &  0.109 &  0.136 &  0.363(-1) 
\\ 5 &  0.495 &  0.205(1) & 0.128(2) &  0.111(2) &  0.400(3) &  0.264(1) & 0.166(1) & 0.123(2) &  0.510(2) 
\\ 6 &  0.327(-1) & 0.361 &  0.213(-1) & 0.422 &  0.302(-1) & 0.223(1) & 0.468(-1) & 0.366(-1) & 0.133(-1) 
\\ 7 &  0.500 &  0.286(1) & 0.591(2) &  0.965(1) & 0.166(1) & 0.415(1) & 0.106(3) &  0.146(1) & 0.291(2) 
\\ 8 &  0.331 &  0.138(1) & 0.135(2) &  0.719(1) & 0.735(1) & 0.193(1) & 0.879 &  0.695(2) &  0.302(2) 
\\ 9 &  0.965(-1) & 0.538 &  0.464(1) & 0.192(1) & 0.306(2) &  0.706 &  0.175(2) &  0.302(2) &  0.140(3) 
\\ \hline
\multicolumn{7}{c}{$E_1=0.91\text{Ry}$, $\mbox{}^3S^e, p=-1 $} \\ \hline
   1 &  0.186(1) & 0.408(-1) & 0.669(-2) & 0.170(-1) & 0.571(-2) & 0.158(-1) & 0.287(-2) & 0.160(-2) & 0.258(-3) 
\\ 2 &  0.267 &  0.277(2) &  0.983(-1) & 0.972 &  0.415 &  0.398(1) & 0.188(1) & 0.309(1) & 0.174(1) 
\\ 3 &  0.376(1) & 0.862(1) & 0.175(4) & 0.264(3) &  0.235(3) &  0.266(3) &  0.115(3) &  0.886(2) &  0.392(3) 
\\ 4 &  0.360(-1) & 0.324 &  0.996 &  0.345(1) & 0.563 &  0.698 &  0.542(-1) & 0.574(-1) & 0.483(-1) 
\\ 5 &  0.108(1) & 0.122(2) &  0.785(2) &  0.498(2) &  0.796(3) &  0.440(2) &  0.186(3) &  0.273(2) & 0.202(2) 
\\ 6 &  0.337(-1) & 0.133(1) & 0.101(1) & 0.698 &  0.500 &  0.707(1) & 0.227(-2) & 0.105 &  0.151 
\\ 7 &  0.554 &  0.553(2) &  0.382(2) &  0.482(1) & 0.185(3) &  0.193 &  0.233(3) &  0.694(2) &  0.754(2) 
\\ 8 &  0.189 &  0.544(2) &  0.177(2) &  0.303(1) & 0.164(2) &  0.557(1) & 0.416(2) &  0.106(3) &  0.980(2) 
\\ 9 &  0.314(-1) & 0.307(2) &  0.784(2) &  0.253(1) & 0.121(2) &  0.799(1) & 0.453(2) &  0.981(2)&  0.26(3) 
\\ \hline
\end{tabular}
\end{table}

\end{document}